# SEPARABILITY AND CORRELATIONS IN COMPOSITE STATES BASED ON ENTROPY METHODS

**A. K. Rajagopal and R. W. Rendell**
**Naval Research Laboratory, Washington DC 20375 - 5320**

## ABSTRACT

This work is an enquiry into the circumstances under which entropy methods can give an answer to the questions of both quantum separability and classical correlations of a composite state. Several entropy functionals are employed to examine the entanglement and correlation properties guided by the corresponding calculations of concurrence. It is shown that the entropy difference between that of the composite and its marginal density matrices may be of arbitrary sign except under special circumstances when conditional probability can be defined appropriately. This ambiguity is a consequence of the fact that the overlap matrix elements of the eigenstates of the composite density matrix with those of its marginal density matrices also play important roles in the definitions of probabilities and the associated entropies, along with their respective eigenvalues. The general results are illustrated using pure and mixed state density matrices of two-qubit systems. Two classes of density matrices are found for which the conditional probability can defined: (1) density matrices with commuting decompositions and (2) those which are decohered in the representation where the density matrices of the marginals are diagonal. The first class of states encompass those whose separability is currently understood as due to particular symmetries of the states. The second are a new class of states which are expected to be useful for understanding separability. Examples of entropy functionals of these decohered states including the crucial isospectral case are discussed.

## I. INTRODUCTION

The theory of quantum entanglement has occupied a central place in modern research because of its promise of enormous utility in quantum computing, crypotography, etc. A major thrust of current research is to find a quantitative measure of entanglement for general states. Approaches to this question based on the eigenvalue spectra of the system density matrices such as entropy methods, have given necessary but not sufficient conditions for particular states. However, it is not known for what classes of states entropy conditions apply. Recently important questions have been raised [1, 2] concerning the ability of entropy methods to decide on the question of separability of a composite state. In particular the case of a Werner state of two-qubits $\rho_p = p|\Psi\rangle\langle\Psi| + (1-p)\, I/4\, (0 \le p \le 1)$ and $|\Psi\rangle = (|00\rangle + |11\rangle)/\sqrt{2}$ has been cited [1] to point out that it is separable by partial transpose criterion if and only if $p \le 1/3$ whereas the von Neumann conditional entropy criterion, $S_1(A|B) \equiv S_1(A,B) - S_1(A) \ge 0$ gives the condition of separability for $p \le 0.747\cdots$. Abe and Rajagopal [3] have also pointed out this result as well as the condition for separability based on Bell inequality which gives $p \le 1/\sqrt{2} \cong 0.7071\cdots$, and obtained the necessary condition for separability $p \le 1/3$ by employing the conditional Tsallis entropy condition. This condition namely, $\lim_{q\to\infty} S_q(A|B) \ge 0$, $S_q(A|B) = \{(S_q(A,B) - S_q(A))/(1+(1-q)S_q(A))\}$, with the Tsallis entropy defined as $S_q(\rho) = \{(Tr\rho^q - 1)/(1-q)\}$, was derived by them under *the assumption of the existence of conditional probability*. It should be noted that for q=1, the Tsallis entropy as well as the q-conditional entropy become the corresponding von Neumann versions of entropy, $S_1(\rho) = -\{Tr\rho \ln \rho\}$ and conditional entropy. In an extension of this work, Abe [4] has shown that the Tsallis entropy condition also gives the correct separability criterion for generalized Werner states of N qudits. (See [1] for a derivation of the same result by disorder criterion and references therein to other methods of arriving at the same result.) The above quoted entropy criteria are necessary but not sufficient and the equality signs in them give the demarcation of separability from the entangled regions. Nielsen and Kempe [1] (NK) further provided a crucial isospectral example where two density matrices have the same spectra both globally and locally, one of which is entangled and the other is separable from the partial transpose condition or equivalently the positivity of concurrence. For this isospectral case, the Tsallis entropies are all equal for all q and so the conditional entropies are zero for all q. NK conclude that separability criteria based solely on eigenvalues of the composite density matrix and its marginal density matrices can never work. The same conclusion is advocated in [2] who also claim that the use of Tsallis conditional entropy used in [3, 4] is wrong. Later in this paper, we will clarify the derivation given in [3, 4] and show that it is indeed correct.

From the above examples it is seen that entropy criteria sometimes succeed and sometimes fail in identifying the separability of states. The purpose of this paper is to seek under what conditions entropy methods can give an answer to the questions of both quantum separability and classical correlations of a composite state and identify the reasons for it. It is found that separable states have a fixed sign for the entropy difference whenever the conditional probability could be defined. Furthermore, we find two classes of states for which this can be done. The first class of states, those with commuting decompositions, encompass most of the known examples for which the understanding of the separability conditions took advantage of particular symmetries of the states [3 – 7]. The second class of states, which are decohered in the representation where the density matrices of the marginals are diagonal, are a new class with properties that are expected to be useful in studying entanglement. These decohered states lead us to consider new entropy functionals.

To this end, we first consider the most general form of a bipartite density matrix $\rho(A,B)$, second, its special representation, and third, a further special form for it. The special representation employed here is based on the diagonal representation of the marginal density matrices of $\rho(A,B)$, $\rho(A) = Tr_B\rho(A,B)$ and $\rho(B) = Tr_A\rho(A,B)$. This representation will be designated the $\{\alpha,\beta\}$ representation. In this representation, $\rho(A,B)$ is not diagonal, in general. However, this representation has a crucial property with respect to decoherence. If $\rho(A,B)$ were to decohere in this representation, so that its off-diagonal elements vanished resulting in a decohered density matrix, $\rho_d(A,B)$, it is found that the marginal density matrices of both $\rho(A,B)$ and $\rho_d(A,B)$ are the same. In the discussion of entropy methods for describing the entanglement issues, this representation also makes it transparent that the overlap matrix elements of the composite states with the product states of the marginal density matrices play important roles. We show that only when the eigenstates conspire suitably, the conditional probabilities can be unambiguously defined, thereby allowing entropy inequalities to be established. This is found to happen for two classes of states. One is the class of decohered states $\rho_d(A,B)$. The second class are states which have commuting operators in the subsystems of their decompositions. The corresponding forms for the entropies for each of these classes will be investigated to yield information about the separability and correlations inherent in them**.** It is important to point out that an entropy functional based on the decohered density matrix is shown here to distinguish the entangled and the separable states in the isospectral example given in [1].

In order to illustrate the relationship of the properties of the entropies employed in examining the entanglement status, we use as a guide the "concurrence" measure [8, 9] which is a necessary and sufficient condition for entanglement of two-qubit density matrices. Thus it is shown that the eigenstates corresponding to the eigenvalues must also be included in deducing the entanglement status of a composite system. These results are valid both for pure and mixed state density matrices. In the case of a general two-qubit pure state density matrix, this formulation is shown to give a necessary and sufficient condition for the separability. Several cases of bipartite mixed state density matrices are used to illustrate our formulation. It should be emphasized that the conditions obtained for the mixed states however are only necessary but not sufficient.

This paper is divided into four sections. In Sec.II, the special representations in which the marginal density matrices are diagonal and the ensuing properties are developed. We employ these to represent the known concepts of entanglement and correlations. Section III contains the main results of this work presented in the form of four theorems. We illustrate these results with specific examples of bipartite density matrices to elucidate the theorems. A complete account of a pure state of general two-qubit system is also included in this section. In the final Sec.IV, discussion and concluding remarks are given along with the introduction of entropy functionals based on the decohered states. In Appendix I, details are given of the general two-qubit pure state density matrix in Sec.III.

## II. PRELIMINARIES IN TERMS OF SPECIAL REPRESENTATIONS

Consider a bipartite state of a composite system described by the density matrix $\rho(A,B)$ whose marginal density matrices are $Tr_B\rho(A,B)=\rho(A)$ and $Tr_A\rho(A,B)=\sigma(B)$. These density matrices being Hermitian, trace-class operators have their own diagonal representations in terms of orthonormal and complete states (here we employ discrete, finite number of states for simplicity, as in the cases of qudit systems) in their respective spaces:

$$\rho(A,B)=\sum_{\Gamma}|\Gamma\rangle P(\Gamma)\langle\Gamma|,\ \ \rho(A)=\sum_{\alpha}|\alpha\rangle p(\alpha)\langle\alpha|,\quad \sigma(B)=\sum_{\beta}|\beta\rangle q(\beta)\langle\beta| \tag{1}$$

$$\langle\Gamma|\Gamma'\rangle=\delta_{\Gamma,\Gamma'},\quad \sum_{\Gamma}|\Gamma\rangle\langle\Gamma|=I(A,B)\equiv I(A)\otimes I(B);$$
$$\langle\alpha|\alpha'\rangle=\delta_{\alpha,\alpha'},\ \sum_{\alpha}|\alpha\rangle\langle\alpha|=I(A),\quad \langle\beta|\beta'\rangle\delta_{\beta,\beta'},\ \ \sum_{\beta}|\beta\rangle\langle\beta|=I(B) \tag{2}$$

Here I stands for identity operators in the spaces specified. Here it should be noted that $|\Gamma\rangle$ represents the composite state of the (A,B) system in its most general form. In general, it is an entangled state. Also, since these are all density matrices, $P(\Gamma), p(\alpha)$, and $q(\beta)$ are the corresponding probabilities and hence are positive, taking the values between 0 and 1. The marginal density matrices may also be expressed in the following alternate forms:

$$\rho(A)=Tr_B\rho(A,B)=\sum_{\Gamma}\sum_{\beta}\langle\beta|\Gamma\rangle P(\Gamma)\langle\Gamma|\beta\rangle\equiv\sum_{\alpha}|\alpha\rangle p(\alpha)\langle\alpha|, \tag{3}$$

$$\sigma(B)=Tr_A\rho(A,B)=\sum_{\Gamma}\sum_{\alpha}\langle\alpha|\Gamma\rangle P(\Gamma)\langle\Gamma|\alpha\rangle\equiv\sum_{\beta}|\beta\rangle q(\beta)\langle\beta|, \tag{4}$$

so that we have the following relations between the eigenvalues of the marginal density matrices and those of the composite density matrix

$$p(\alpha)=\sum_{\Gamma}\sum_{\beta}|\langle\alpha,\beta|\Gamma\rangle|^2P(\Gamma) \tag{5}$$

and $$q(\beta)=\sum_{\Gamma}\sum_{\alpha}|\langle\alpha,\beta|\Gamma\rangle|^2P(\Gamma). \tag{6}$$

Here we have introduced the notation $|\alpha,\beta\rangle=|\alpha\rangle|\beta\rangle$. It is important to note that $p(\alpha), q(\beta)$ are *not marginal probabilities* of the composite probability $P(\Gamma)$ in view of the appearance of the overlap matrix element in eqs.(5, 6).

By using the completeness relations we may also express the density matrix of the composite system in terms of the states of A and B systems in which representation $\rho(A,B)$ is not necessarily diagonal:

$$\rho(A,B)\equiv\sum_{\alpha,\beta}\sum_{\alpha',\beta'}\sum_{\Gamma}|\alpha,\beta\rangle\langle\alpha,\beta|\Gamma\rangle P(\Gamma)\langle\Gamma|\alpha',\beta'\rangle\langle\alpha',\beta'|. \tag{7}$$

Working out the marginal density matrices from this expression and comparing it with eqs.(3, 4) and (5, 6), we obtain the following expressions:

$$\sum_{\Gamma}\sum_{\beta}\langle\alpha,\beta|\Gamma\rangle P(\Gamma)\langle\Gamma|\alpha',\beta\rangle=p(\alpha)\delta_{\alpha,\alpha'} \tag{8}$$

and $$\sum_{\Gamma}\sum_{\alpha}\langle\alpha,\beta|\Gamma\rangle P(\Gamma)\langle\Gamma|\alpha,\beta'\rangle = q(\beta)\delta_{\beta,\beta'} \ . \qquad (9)$$

There has been some recent suggestions to use another quantity called the mutual entropy, which is used in classical contexts to examine the classical correlations among the variables, to examine the quantum entanglement/correlations [10 - 12]. It is defined as

$$S_1(A:B) = S_1(A) + S_1(B) - S_1(A,B) \qquad (10)$$

Here $S_1$ stands for the von Neumann entropy. If $\rho(A,B) = \rho(A)\otimes\sigma(B)$, then $S_1(A:B)\equiv 0$ and conversely, if $S_1(A:B)\equiv 0$, then $\rho(A,B) = \rho(A)\otimes\sigma(B)$. Such a composite density matrix has no correlation between its subsystems. This result is also true for classical systems. The concept of mutual entropy in the Tsallis theory is not expressible in a neat form in view of the fact that the bipartite density matrix is not diagonal in the $\{|\alpha,\beta\rangle\}$ representation (see eq.(7)).

A composite density matrix is said to be quantum-separable if it can be written as convex combinations of individual density matrices of A and B in the form $\rho(A,B) = \sum_j w_j \rho_j(A)\otimes\sigma_j(B), (0\le w_j \le 1), \sum_j w_j = 1$. For classically correlated systems, a similar decomposition holds and hence one often uses the terms "classically correlated" and "quantum separable" interchangeably to describe such systems. For the special case when all the $w_i$'s are zero except for one, this expression reduces to the uncorrelated case where the mutual entropy is zero. In the more general form, the mutual entropy may be nonzero thus indicating the presence of correlations in the system.

In some simple cases, one may express the composite density matrix in the form given above and check its separability. For example, the Werner state quoted above may be written in the form

$$\rho_p = (1-3p)\, I/4 + \frac{p}{2}\left\{\sum_{i=1,3;\varepsilon=\pm}\rho_i^{\varepsilon}(A)\otimes\sigma_i^{\varepsilon}(B) + \sum_{i=2;\varepsilon=\pm}\rho_i^{\varepsilon}(A)\otimes\sigma_i^{-\varepsilon}(B)\right\}$$ where

$\rho_i^{\varepsilon}(A) = \sigma_i^{\varepsilon}(B) = (I_2 + \varepsilon\hat{\sigma}_i)/2$, with $\hat{\sigma}_i$ 's are the standard Pauli 2x2 matrices and $I_2$ is a 2x2 unit matrix. One immediately notices that separability of the Werner state follows if $p\le 1/3$, when the weights are all positive. Note that this result is derived without examining the eigenvalues at all. But this direct method has not yielded a useful measure for the entanglement of general states.

Concurrence [8, 9] is a valid measure of entanglement of two qubits. The system is separable if and only if C(AB)=0 and when C(AB)= 1, it is maximally quantum entangled. The concurrence C(AB) of a density matrix $\hat{\rho}(AB)$ [8, 9] in the computational basis will be used here. It is defined by first constructing the matrix, $\hat{\tilde{\rho}}(AB) = (\hat{\sigma}_2\otimes\hat{\sigma}_2)\,\hat{\rho}^*(AB)(\hat{\sigma}_2\otimes\hat{\sigma}_2)$, where $\hat{\sigma}_2$ is the standard y-component 2x2 Pauli matrix, and $\hat{\rho}^*(AB)$ is the complex conjugate matrix of $\hat{\rho}(AB)$. The concurrence is then given by $C(AB) = \max\{\lambda_1 - \lambda_2 - \lambda_3 - \lambda_4, 0\}$ where $\{\lambda_1, \lambda_2, \lambda_3, \lambda_4\}$ are the square roots of the eigenvalues of the matrix product $\hat{\rho}(AB)\hat{\tilde{\rho}}(AB)$ arranged in decreasing order. It was also shown to be a symmetry dependent measure, e.g. permutation symmetry [13, 14]. In the Werner example, C(AB)= max{(3p-1)/2, 0} and this is in agreement with the result quoted above which was obtained from the separable decomposition of the density matrix. Recall that the Tsallis entropy criterion [3] gives the same correct p values for the entanglement status of the Werner state. We will presently demonstrate that this criterion depends on the eigenvalues of the composite density matrix and its eigenfunctions in a special way. In view of this result, expressions (1) will be used here to examine to what extent one can separate the quantum entanglement from the classical correlations contained in a composite density matrix.

We now show that the eigenvalues defined above, $\{\lambda_1, \lambda_2, \lambda_3, \lambda_4\}$, may be expressed as the eigenvalues of $\hat{\rho}(AB)\hat{\tilde{\rho}}(AB) = \sum_{\Gamma,\Gamma'} |\Gamma\rangle C(\Gamma,\Gamma')\langle\Gamma'|$, with $C(\Gamma,\Gamma') = \sum_{\Gamma_1} P(\Gamma)\langle\Gamma|\hat{\sigma}_2 \otimes \hat{\sigma}_2|\Gamma_1^*\rangle P(\Gamma_1)\langle\Gamma_1^*|\hat{\sigma}_2 \otimes \hat{\sigma}_2|\Gamma\rangle$. Here the * denotes complex conjugate. Thus the eigenvalues involved in the computation of concurrence are related to both the eigenvalues of the density matrix of (AB) and an appropriate set of its matrix elements as indicated in the above expression. This already gives a hint that the eigenfunctions generally also play an important role in determining the entanglement status of the system. This will be discussed in more detail in the following paragraphs.

Special cases of eq.(1) will now be discussed depending on the context defining the composite state. We state the central results of this paper in the form of four theorems. We will also give some illustrative examples accompanying theorems A, B and D which are summarized in Table I. Possible entropy functionals based on Theorem C are discussed in Sec. IV and an example of this, the quantum deficit, is also included in Table I.

## III. MAIN RESULTS

**Theorem A**: The general composite mixed state density matrix and its marginal density matrices do not have a fixed sign for the entropy differences, $\{S_1(A,B) - S_1(A)\}$ or $\{S_1(A,B) - S_1(B)\}$ and they may even be zero.

Theorem A follows because, by the above considerations, we have

$$S_1(A,B) = -\sum_{\Gamma} P(\Gamma)\ln P(\Gamma) = -\sum_{\Gamma}\sum_{\alpha,\beta} |\langle\alpha,\beta|\Gamma\rangle|^2 P(\Gamma)\ln P(\Gamma) \tag{11}$$

and $$S_1(A) = -\sum_{\alpha} p(\alpha)\ln p(\alpha) = -\sum_{\Gamma}\sum_{\alpha,\beta} |\langle\alpha,\beta|\Gamma\rangle|^2 P(\Gamma)\ln p(\alpha) \tag{12}$$

where eq.(6) was used. Similarly $S_1(B)$ may be written down. Each of the entropies defined in eqs.(11, 12) is positive. But the difference between eqs.(11, 12) is $S_1(A,B) - S_1(A) = -\sum_{\Gamma}\sum_{\alpha,\beta} |\langle\alpha,\beta|\Gamma\rangle|^2 P(\Gamma)\ln(P(\Gamma)/p(\alpha))$ and a similar expression for the other difference. Since $p(\alpha), q(\beta)$ are not marginal probabilities of the composite probability $P(\Gamma)$, the ratio $P(\Gamma)/p(\alpha)$ or $P(\Gamma)/q(\beta)$ may be greater than unity and hence no conclusion can be drawn about the sign of $\{S_1(A,B) - S_1(A)\}$ or $\{S_1(A,B) - S_1(B)\}$. Only when *the notion of conditional probability can be defined,* such a ratio will lie between 0 and 1, and the above differences of entropies become "conditional entropies" having only positive sign. Under special circumstances, they can also be zero.

The relation between the sign of the entropy difference to entanglement must be inferred by other considerations to be discussed later in this section.

It is interesting to note that in the Werner case, the eigenvalues of the composite state are $P_W(\Gamma) = \left(\frac{3p+1}{4}, \frac{1-p}{4}, \frac{1-p}{4}, \frac{1-p}{4}\right)$, and the two marginal density matrices are the same and their eigenvalues are $p_W(\alpha) = \left(\frac{1}{2}, \frac{1}{2}\right) = q_W(\beta)$. Therefore the ratio $P_W/q_W$ will qualify to be conditional probability if $((3p+1)/4)/(1/2) \le 1$ or for $p \le 1/3$. This is the same condition for separability derived earlier.

Three entangled mixed states which have a negative, positive, and zero entropy difference are now given to illustrate Theorem A. These examples, as well as the ones used to illustrate the other theorems, are physically motivated and are all marginals of three-qubit pure state density matrices [13] which are related to GHZ and W classes of tripartite states. The results for all the examples are collected together in Table I.

(E1) Mixed State

$$\rho_1(AB) = \frac{1}{6}\left((-2|10\rangle + |01\rangle)(-2\langle 10| + \langle 01|) + |00\rangle\langle 00|\right)$$
$$\rho_1(A) = \frac{1}{6}\left(4|1\rangle\langle 1| + 2|0\rangle\langle 0|\right), \quad \sigma_1(B) = \frac{1}{6}\left(|1\rangle\langle 1| + 5|0\rangle\langle 0|\right) \tag{13}$$

The marginal density matrices are already diagonal and thus saves one step in our procedure. This is the situation for all the examples chosen here to illustrate the various theorems. The eigenvalues of the density matrices in eq.(13) are $P_1(\Gamma) = \left(0, 0, \frac{1}{6}, \frac{5}{6}\right)$ and $p_1(\alpha) = \left(\frac{1}{3}, \frac{2}{3}\right), q_1(\beta) = \left(\frac{1}{6}, \frac{5}{6}\right)$.

The entropy differences are $S_1(AB) - S_1(A) = \frac{5}{6}\ln\left(\frac{4}{5}\right) < 0$ and $S_1(AB) - S_1(B) = 0$.

(E2) Mixed State

$$\rho_2(BC) = \frac{1}{6}\left(\left(|10\rangle + |01\rangle\right)\left(\langle 10| + \langle 01|\right) + 4|00\rangle\langle 00|\right),$$
$$\rho_2(B) = \frac{1}{6}\left(|1\rangle\langle 1| + 5|0\rangle\langle 0|\right) = \sigma_2(C) \qquad (14)$$

The eigenvalues are $P_2(\Gamma) = \left(0, 0, \frac{2}{6}, \frac{4}{6}\right)$ and $p_2(\alpha) = \left(\frac{1}{6}, \frac{5}{6}\right) = q_2(\beta)$.

The entropy differences are equal and $S_1(BC) - S_1(B) = \frac{5}{6}\ln\left(\frac{5}{4}\right) > 0$ .

(E3) Mixed State

$$\rho_3(AB) = \frac{1}{3}\left(\left(|10\rangle + |01\rangle\right)\left(\langle 10| + \langle 01|\right) + |00\rangle\langle 00|\right),$$
$$\rho_3(A) = \frac{1}{3}\left(|1\rangle\langle 1| + 2|0\rangle\langle 0|\right) == \sigma_3(B) \qquad (15)$$

And their eigenvalues are $P_3(\Gamma) = \left(0, 0, \frac{1}{3}, \frac{2}{3}\right)$ and $p_3(\alpha) = \left(\frac{1}{3}, \frac{2}{3}\right) = q_3(\beta)$.

The entropy differences are zero, $S_q(AB) - S_q(A) = 0$ for all q.

Examples (E1) - (E3) are all entangled states as shown by the concurrence values in Table I.

The property of the mutual entropy that $S_1(A:B) \geq 0$ is a known result but a proof of this in terms of the $\{\alpha, \beta\}$ basis is illuminating because

$$S_1(A:B) = \sum_{\Gamma}\sum_{\alpha,\beta} \left|\langle \alpha, \beta | \Gamma \rangle\right|^2 P(\Gamma) \ln\left(P(\Gamma)/p(\alpha)q(\beta)\right) \geq 0. \qquad (16)$$

The inequality follows from the property of the logarithm: $\ln X \geq 1 - (1/X)$, for positive X, and the various normalizations shown in eq.(2). The equality sign is obtained when the composite system is both unentangled and uncorrelated i.e., when $P(\Gamma) = p(\alpha)q(\beta)$ or more generally when $\rho(A,B) = \rho(A) \otimes \sigma(B)$. This result is both necessary and sufficient.

Although the sign of the entropy differences is not fixed in general, the following three theorems identify situations where the conditional probabilities can be defined and the sign becomes fixed.

**Theorem B:** The entropy differences are in general less than or equal to zero for an entangled pure state of two qubits.

This follows by giving a complete account of a pure state density matrix. In Appendix A, we give the general formulas needed in developing the discussion of the pure state density matrix of a two-qubit state. The representations that diagonalize the marginal density matrices given in eqs.(A11, 12) are

$$\rho(A) = |\alpha_1\rangle p(\alpha_1)\langle\alpha_1| + |\alpha_2\rangle p(\alpha_2)\langle\alpha_2|,$$
$$p(\alpha_1), p(\alpha_2) = \left(1 \pm |\vec{s}(A)|\right)/2. \tag{17}$$

$$\sigma(B) = |\beta_1\rangle q(\beta_1)\langle\beta_1| + |\beta_2\rangle q(\beta_2)\langle\beta_2|,$$
$$q(\beta_1), q(\beta_2) = \left(1 \pm |\vec{s}(B)|\right)/2. \tag{18}$$

The corresponding eigenvectors in terms of the familiar computational basis are given by

$$|\alpha_1\rangle = A_+|1\rangle + e^{i\phi(A)}A_-|0\rangle, \quad |\alpha_2\rangle = -e^{-i\phi(A)}A_-|1\rangle + A_+|0\rangle,$$
$$A_\pm = \left(\left(|\vec{s}(A)| \pm s_3(A)\right)/2|\vec{s}(A)|\right)^{1/2}; e^{i\phi(A)} = \frac{s_1(A) + is_2(A)}{\sqrt{s_1^2(A) + s_2^2(A)}}. \tag{19}$$

with similar expressions for the B-system. One may then compute the system density matrix in terms of these. Since this system is a pure state, we have

$$S_1(A, B) = 0, \; S_1(A|B) = -S_1(B), \quad S_1(A:B) = 2S_1(A) \tag{20}$$

The entropies of A and B are equal because of eq.(A14) and given by

$$0 \le S_1(A) = S_1(B) = -p(\alpha_1)\ln p(\alpha_1) - p(\alpha_2)\ln p(\alpha_2) \le 1. \tag{21}$$

It should be noted that eq.(A14) expresses the magnitudes of the spin vectors of A and B in terms of the known concurrence [8, 9, 18] for the pure state given by eq.(A2). The following interesting cases are evident from eqs.(A11, 12):
Case (A): The reduced density matrices are also pure states if the vectors $\vec{s}(A)$ and $\vec{s}(B)$ are unit vectors, when the equality sign in eq.(A13) is satisfied. The corresponding entropies are then zero. In this case we obtain the pure state to be **separable**. It turns out that this is also sufficient for the separability of the pure state.
Case (B): The reduced matrices are maximally chaotic mixed states if either of the two spin vectors is zero, and hence the pure state is **maximally entangled**. In this case, the entropy of A is ln2. The conditional entropy is negative, –ln2, and the mutual entropy is 2ln2.
Case (C): It follows from Cases A and B that the inequality in eq.(A13) shows that for all other polarizations of the qubits, $1/2 \le Tr_A \rho^2(A) < 1$ and the pure state is entangled. This also implies that the conditional entropy is negative.

Thus we may conclude that the entropy method works for the discussion of entanglement of two-qubit pure states. The conditional entropy is negative if and only if it is entangled.

An example of an entangled pure state density matrix is (E4) given below.

(E4) Pure State

$$\rho_4(BC)=\frac{1}{2}\left((|10\rangle-|01\rangle)(\langle 10|-\langle 01|)\right),$$
$$\rho_4(B)=\frac{1}{2}\left(|1\rangle\langle 1|+|0\rangle\langle 0|\right)=\sigma_4(C) \tag{22}$$

The eigenvalues are $P_4(\Gamma)=(0,0,0,1)$ and $p_4(\alpha)=\left(\frac{1}{2},\frac{1}{2}\right)=q_4(\beta)$.

The entropy differences are equal and $S_1(BC)-S_1(B)=-\ln 2<0$. This is a pure state density matrix and, as shown above, the negative entropy difference implies it is entangled. This agrees with the result for the concurrence, C(AB)=1.

**Theorem C**: If the eigenfunctions conspire in such a way that one may express the composite density matrix in the"decohered" form

$$\rho_d(A,B)=\sum_{\alpha,\beta}|\alpha,\beta\rangle P_d(\alpha,\beta)\langle\alpha,\beta| \tag{23}$$

where $$P_d(\alpha,\beta)=\sum_{\Gamma}|\langle\alpha,\beta|\Gamma\rangle|^2 P(\Gamma) \tag{24}$$

then $p(\alpha)$, $q(\beta)$ are indeed marginal probabilities of the composite probability $P_d(\alpha,\beta)$, and $S_{1d}(A,B)\geq S_1(A),S_1(B)$. Also, the entropy of a general $\rho(A,B)$ is less than or equal to that described by eq.(23): $S_1(A,B)\leq S_{1d}(A,B)$. Equality is obtained when $\rho(A,B)=\rho_d(A,B)$ and this can happen if $\rho(A,B)$ commutes with $\rho(A)\otimes I(B)$ and $I(A)\otimes\sigma(B)$.

We first discuss the significance of this diagonal form of the density matrix in the $(\alpha,\beta)$ representation which diagonalizes the marginal density matrices. Equation (23) in comparison with eq.(7) implies that $\sum_{\Gamma}\langle\alpha,\beta|\Gamma\rangle P(\Gamma)\langle\Gamma|\alpha',\beta'\rangle=P_d(\alpha,\beta)\delta_{\alpha,\alpha'}\delta_{\beta,\beta'}$ and eqs.(8, 9) would then follow. This can happen if and only if the system density matrix $\rho(A,B)$ commutes with $\rho(A)\otimes I(B)$ and $I(A)\otimes\rho(B)$. In Theorem D, and in several examples given below including the Werner case, such a situation occurs. Equation (23) can arise when there is “decoherence” of the system either due to the effects of environment or due to a measurement process as in [12]. In that case, the off-diagonal elements in eq.(7) are eliminated and only the diagonal elements endure. It should be emphasized that decoherence is basis dependent and $\rho_d(A,B)$ can in general be entangled because it need not be diagonal in the computational basis. For example compare the Werner state in the Bell basis and the computational basis. It is important to note that our choice of the $\{\alpha,\beta\}$ representation is such that the marginal density matrices of the decohered system are the same as those for the original system.

Proof of this theorem is straightforward. In this case we can define conditional probabilities as in the classical case, because $0\leq P_d(\alpha|\beta)=P_d(\alpha,\beta)/q(\beta)\leq 1$ etc. and conditional entropy statements both for Tsallis and von Neumann cases may be deduced [3]. It should be noted that the decohered density matrix given by eq.(23) is in general quantum entangled because it cannot always be expressed as a convex combination of pure state density matrices $\{\rho_i(A)\otimes\sigma_i(B)\}$ with positive weights $w_i$ such that $\sum_i w_i = 1$. If, however we are given the entropy inequality, in view of Theorem A, we cannot conclude that the composite density matrix is of the form eq.(23). Also we cannot deduce the form of the joint probability given in eq.(24).

The last statement in the above theorem is originally due to Klein [15]. He showed that whenever the off-diagonal elements of a density matrix are discarded (due to "random phase approximation" as was then a prevalent concept), the von Neumann entropy increases. A simple proof of this follows from the Kullback – Leibler relation, $Tr\rho_1(\ln\rho_1 - \ln\rho_2) \geq 0$ where $\rho_1$ and $\rho_2$ are any two density matrices in the same space. Taking $\rho_1 = \rho(A,B)$ and $\rho_2 = \rho_d(A,B)$ we have $-S_1(A,B) - Tr\rho(A,B)\ln\rho_d(A,B) \geq 0$ ;but from eq.(23), the result is thus established.

A second class of density matrices for which conditional probability exists is found for commuting operators within each subsystem of local decompositions. A composite density matrix is quantum separable (or classically correlated) if it can be written in the form of a convex combination

$$\rho(A,B) = \sum_j w_j \rho_j(A)\otimes\sigma_j(B), (0 \leq w_j \leq 1), \sum_j w_j = 1. \qquad (25a)$$

This is a local representation in which the weights $w_j$ can be interpreted as classical probabilities. Any entangled state can also be given a local representation in the form called a local pseudo-mixture [16]

$$\rho(A,B) = -t\rho^-(A,B) + (1+t)\rho^+(A,B) \qquad (25b)$$

where each of $\rho^{\pm}(A,B)$ are of the separable form given in eq.(25a) and t is a finite positive number. Thus pseudo-mixtures involve negative coefficients in their local representations and these do not have the properties of classical probabilities. The representation (25b) is not unique, but the minimum value of t represents a valid measure of entanglement [16].

The next theorem relates the separability condition of eq.(25a) to an entropy difference condition.

**Theorem D**: If in a local representation of a composite density matrix, the density matrices of A and B for different j commute in their respective spaces, then $S_1(A,B) \geq S_1(A), S_1(B)$. When they do not commute, the results are as in Theorem A.

Consider first the separable case given by eq.(25a). This equation leads to the following marginal density matrices of A and B:

$$\rho(A) = \sum_j w_j \rho_j(A) \equiv \sum_\alpha |\alpha\rangle p(\alpha)\langle\alpha|$$

with $\quad p(\alpha) = \sum_j w_j \langle\alpha|\rho_j(A)|\alpha\rangle$, and similarly

$$\sigma(B) = \sum_j w_j \sigma_j(B) \equiv \sum_\beta |\beta\rangle q(\beta)\langle\beta| \tag{26a,b}$$

with $\quad \mathrm{q}(\beta) = \sum_j w_j \langle\beta|\sigma_j(B)|\beta\rangle$,

where the states of A, B diagonalize the sum of the weighted density matrices. Also, we have in such a representation that the composite density matrix is not diagonal in the $\{\alpha,\beta\}$ basis

$$\rho(A,B) = \sum_{\alpha,\beta}\sum_{\alpha',\beta'} |\alpha,\beta\rangle P(\alpha,\beta;\alpha',\beta')\langle\alpha',\beta'|,$$
$$P(\alpha,\beta;\alpha',\beta') = \sum_j w_j \langle\alpha|\rho_j(A)|\alpha'\rangle\langle\beta|\sigma_j(B)|\beta'\rangle \tag{27}$$

Note that p, q are the marginal probabilities associated with the composite density matrix as in Theorem A and the results for the corresponding entropies are ambiguous because here again we cannot define conditional probabilities. However, when the density matrices of A and B for different j commute in their respective spaces, or if we can find a P in eq.(27) in the form

$$P(\alpha,\beta;\alpha',\beta') = P_D(\alpha,\beta)\delta_{\alpha,\alpha'}\delta_{\beta,\beta'},$$
$$\text{where } P_D(\alpha,\beta) \equiv \sum_j w_j \langle\alpha|\rho_j(A)|\alpha\rangle\langle\beta|\sigma_j(B)|\beta\rangle \tag{28}$$

then the results of Theorem C follow as before. This is because p, q are now the conditional probabilities associated with the composite probability $P_D(\alpha,\beta)$ defined in eq.(28).

The composite state in Theorem D turns out to be $|\Gamma\rangle = |a,b\rangle$ because one can diagonalize simultaneously the density matrices $\rho_j(A)$ for all j. In this case, one obtains the following simpler representations of the marginal density matrices $\rho_j(A) = \sum_a |a\rangle p_j(a)\langle a|$, $\sigma_j(B) = \sum_b |b\rangle q_j(b)\langle b|$, and hence the composite density matrix takes the diagonal form $\rho_d(A,B) = \sum_{a,b} |a,b\rangle P(a,b)\langle a,b|$ where $P(a,b) = \sum_j w_j p_j(a) q_j(b)$. Then $p(a) = \sum_j w_j p_j(a)$ and $q(b) = \sum_j w_j q_j(b)$ are marginals of P(a,b). Then the conditional probabilities can be defined and thus the inequalities follow.

Again, we see that if the entropy inequalities are obtained for a given composite system, one cannot conclude about the separability of the density matrix.

Under the conditions of Theorem D, eq.(16) takes a simpler form:

$$S_1(A:B) = \sum_{\alpha,\beta} P(\alpha,\beta)\ln\left(P(\alpha,\beta)/p(\alpha)q(\beta)\right) \geq 0\,. \qquad (29)$$

In eq.(25a), which is different from $\rho(A,B) = \rho(A) \otimes \sigma(B)$, the equality sign in eq.(29) implies classical correlation hidden in it as also suspected in [10 - 12]. The convex combination in eq.(25a) contains in it some classical correlation even though it also defines quantum separability.

If the local representation is of the entangled form given by eq.(25b), then the marginal density matrices are of the form $\rho(A) = -t\rho^-(A) + (1-t)\rho^+(A)$,
and $\rho(B) = -t\rho^-(B) + (1-t)\rho^+(B)$, where the expressions for the terms in the right hand side are of the form given in eq.(26a,b). A discussion following the lines of Theorem D above in this case leading to similar conclusions for the entropy differences is a bit more involved requiring further analysis and will be postponed to a later communication.

We will now give two examples of Theorem D which are of the separable form of eq.(25a) which obey the commuting condition. These are also summarized in Table I.

(E5) Mixed State

$$\begin{aligned} &\rho_5(AB) = \frac{1}{2}\left(|00\rangle\langle 00| + |01\rangle\langle 01|\right), \\ &\rho_5(A) = |0\rangle\langle 0|, \quad \sigma_5(B) = \frac{1}{2}\left(|0\rangle\langle 0| + |1\rangle\langle 1|\right) \end{aligned} \qquad (28)$$

Their eigenvalues are $P_5(\Gamma) = \left(0, 0, \frac{1}{2}, \frac{1}{2}\right)$ and $p_5(\alpha) = (0, 1)$, $q_5(\beta) = \left(\frac{1}{2}, \frac{1}{2}\right)$.
The entropy differences are $S_1(AB) - S_1(A) = \ln 2$ and $S_1(AB) - S_1(B) = 0$.

(E6) Mixed State

$$\begin{aligned} &\rho_6(AB) = \frac{1}{2}\left(|11\rangle\langle 11| + |00\rangle\langle 00|\right), \\ &\rho_6(A) = \frac{1}{2}\left(|1\rangle\langle 1| + |0\rangle\langle 0|\right) = \sigma_6(B) \end{aligned} \qquad (29)$$

The eigenvalues of these are $P_6(\Gamma) = \left(0, 0, \frac{1}{2}, \frac{1}{2}\right)$ and $p_6(\alpha) = \left(\frac{1}{2}, \frac{1}{2}\right) = q_6(\beta)$.
$S_q(A,B) - S_q(A) = 0$ for all q.

In the Werner state example [3] given above and elsewhere [4], this commuting situation occurs allowing conditional probabilities to be defined and thus these works are not wrong as alleged by Vollbrecht and Wolf [2]. When one of the weights $w_i$ is not positive, then the above argument fails as happened in the Werner example. Also, as shown under Theorem A, in the Werner state example the conditional probability can be defined only for $p \leq 1/3$.

## IV. DISCUSSION AND CONCLUDING REMARKS

There are only a few classes of density matrices for which computable entanglement measures and separability conditions are known. Aside from the case of two qubits for which the concurrence formula applies, the other known examples take advantage of particular symmetries of the density matrix to obtain entanglement information [7]. Examples are isotropic states in arbitrary dimensions which are invariant under $U \otimes U^*$ where $U^*$ is the complex conjugate of U in some basis [5] and generalized Werner states of N-qubits [6] which are invariant under all unitary transformations of the form $U \otimes U$. Both of these examples fall under the commuting condition of Theorem D and it can be seen from this point of view why the separability for these classes of density matrices are expected to be tractable. The commuting condition allowed conditional probabilities to be defined and this ingredient is important in defining separability conditions. Thus the results of this paper are pertinent to most of the known classes of density matrices for which entanglement and separability are understood. In Theorem C, we had identified a new second class of density matrices, the decohered density matrices in the $\{\alpha, \beta\}$ basis, for which conditional probabilities can also be defined. The decohered density matrices could thus be another candidate for a class whose entanglement measure and separability conditions may also turn out to be tractable. This class of density matrices therefore deserves further study. In particular, entropy functionals based on the decohered density matrices can be expected to be useful in understanding separability. We give one example of this, which we will call the quantum deficit, to illustrate its features within the examples. However, construction of an entanglement measure based on decohered density matrices awaits further study.

Since $\rho_d(A,B)$ is the classically correlated version of the density matrix of the actual system, we may define the difference between the von Neumann entropy of the system and that of the decohered state as the *quantum deficit*,

$$D(A,B) = S_{1d}(A,B) - S_1(A,B) \geq 0. \tag{30}$$

We also have another important inequality,

$$D(AB) \leq S_1(A:B). \tag{31}$$

The proof of this inequality follows from the fact that

$$D(AB) - S_1(A:B) = S_d(A,B) - S_1(A) - S_1(B), \tag{32}$$

which then is shown to be negative definite following Kullback-Leibler relation. The quantum deficit serves as a measure of the quantum entanglement over and above the classical correlation.

The quantum deficit is different from the *quantum discord* that Oliver and Zurek [12, 17] introduced in that we employ the special states that diagonalize the marginal density matrices whereas the discord does not. The quantum deficit uses a decohered density matrix which maintains the same information contained in the marginal states A, B. This choice removes as much of the ambiguity as possible in comparing correlation and entanglement contributions even after environment and/or measurement effects have taken place in the decoherent process. In the examples given below, the quantum deficit is seen to track closely with the concurrence.

As an example of the utility of the concept of quantum deficit, consider the isospectral example given in [1]. The entangled density matrix is given by $\rho_E(A,B) = \{|11\rangle\langle 11| + |10\rangle\langle 10| + |01\rangle\langle 01| + |10\rangle\langle 01| + |01\rangle\langle 10|\}/3$. This has zero entropy difference, quantum deficit $D_E = (2/3)\ln 2$, and concurrence $C_E = 2/3$. For its separable isospectral counterpart, $\rho_S(A,B) = \{|11\rangle\langle 11| + 2|00\rangle\langle 00|\}/3$, the entropy difference, quantum deficit, and concurrence are all zero. They both have the same set of entropies for both the composite and their marginal density matrices, and therefore their mutual entropies are the same, $(3\ln 3 - 2\ln 2)/3$. Thus the quantum deficit distinguishes the isospectral density matrices whereas the mutual entropy and the entropy difference do not. This is because the decohered entangled state is a convex combination of product form and not given by a single product, $\rho(A)\otimes\sigma(B)$. Although $\rho_E(A,B)$ and $\rho_S(A,B)$ have the same spectra, the corresponding decohered density matrices have different spectra while preserving the reduced density matrices for A and B.

In the Werner state example, the decohered density matrix is found to be $\rho_{pd}(AB) = \left(\frac{1+p}{4}\right)(|11\rangle\langle 11| + |00\rangle\langle 00|) + \left(\frac{1-p}{4}\right)(|10\rangle\langle 10| + |01\rangle\langle 01|)$. And so the quantum deficit is found to be $D_{Wp}(AB) = \left(\frac{1+3p}{4}\right)\ln\left(\frac{1+3p}{4}\right) + \left(\frac{1-p}{4}\right)\ln\left(\frac{1-p}{4}\right) - \left(\frac{1+p}{2}\right)\ln\left(\frac{1+p}{4}\right)$. In Fig.1, we display the behavior of concurrence, mutual entropy, and quantum deficit as a function of the parameter p of the Werner state. This clearly shows the presence of correlations and entanglement in this important example. The Werner state is a special example for which every observation made in this work is verified. In Fig. 1, concurrence, C (full line), mutual entropy scaled by ln2, S (dotted curve), and quantum deficit scaled by ln2, D (dashed curve), are all expressed as a function of the parameter, p, which characterizes the Werner state. The scaling of mutual entropy and the quantum deficit was so as to reflect the inequality $D(AB) \le S_1(A:B)$. Thus the curve for S always lies above that of D. The Werner state is classically correlated (in the sense of eq.(28a)) for $p \le 1/3$ and entangled otherwise; for p=0, it is uncorrelated while for p=1, it is entirely quantum entangled. The scaled S and D are zero for p=0, and for p=1, S=2 and D=1. But for $p \le 1/3$, there are classical correlations remaining because S and D are both finite in this region, with the curve for S lying above the D curve. Also, S never crosses the concurrence line while D lies above for p less than about 0.5 and thereafter it lies below it, approaching 1 as p=1. Thus D/ln2 approximately tracks the quantum entanglement as can be discerned from Fig.1.

Such tracking is also seen in the six examples given to illustrate the various theorems. From Table 1, we observe consistent tracking between the concurrence values and the corresponding scaled quantum deficit, $D/\ln 2$, associated with known composite states. All these results are consequences of the four theorems derived in this paper concerning various forms of the density matrix of a composite system. Of these examples, (E4, E5, E6) are noteworthy. (E4) represents a fully quantum entangled Bell-state and so $D_4(BC)/\ln 2 = 1$ represents the full quantum entanglement as indicated also by $C_4(BC) = 1$. (E5) represents a completely uncorrelated state both classically and quantum mechanically and hence $S_1(A:B), D_5(AB)$, and $C_5(AB)$ are all zero. (E6) displays only classical correlations (being of the form eq.(25a)) with $S_1(A:B)/\ln 2 = 1, D_6(AB) = 0 = C_6(AB)$. The examples (E1, E2) exhibit opposite entropy differences but they are both entangled. (E3) with zero entropy difference is both quantum entangled and classically correlated.

From the above observations, it is tempting to conjecture that there may be a variational principle determining D which will provide an entanglement measure. For example, one may seek the minimum D among all $\rho_d(A,B)$ generated by all transformations of the form $U_1 \otimes U_2$ on $\rho(A,B)$ which give the same marginals. This may be expressed in the form $D = \min\limits_{U_1 \otimes U_2} \{S_d - S | \rho_d(A,B)\}$.

There are three intertwining strands of ideas developed in this paper. The first is the role of conditional probability in determining the sign of the entropy differences of the composite and its marginal density matrices. The second is the importance of the overlap matrix elements of eigenstates of the composite density matrix with those of the marginals. And the third are the decohered density matrices in the representation in which the marginal density matrices are diagonal. These are the bases for the four theorems given in the paper. Together they give insight into the correlations and entanglement of composite density matrices from the entropy considerations.

In summary, we have shown in this paper that the eigenvalues of the composite density matrix and those of its marginal density matrices along with their overlap matrix elements of the respective eigenvectors determine the correlation and entanglement properties of the system. This is accomplished by choosing to study the system in the representation in which the marginal density matrices are diagonal. We present four theorems to elucidate the advantages of this representation and to identify classes of states which have fixed sign for the entropy differences and relation to quantum separability. Explicit computations are presented for two-qubit systems described by (a) Werner state, (b) a general pure state density matrix, and (c) a class of two-qubit mixed states arising out of pure state density matrices of three-qubit states which serve as illustrations of this approach. Thus the inclusion of overlap matrix elements along with the eigenvalues of the density matrix and their marginal density matrices yield information regarding the correlations and entanglement residing in them. This understanding modifies the conclusions reached in [1, 2].

**ACKNOWLEDGEMENT:** Both the authors are supported in part by the Office of Naval Research.

## APPENDIX A
## ON ENTANGLEMENT OF A PURE STATE OF TWO-QUBITS

A general two-qubit pure state density matrix is wriiten in the form

$$\rho(A,B) = |\Psi\rangle\langle\Psi| \tag{A1}$$

where the two-qubit $(2 \otimes 2)$ state is given by

$$|\Psi\rangle = a_{11}|1,1\rangle + a_{10}|1,0\rangle + a_{01}|0,1\rangle + a_{00}|0,0\rangle \tag{A2}$$

with the normalization condition

$$|a_{11}|^2 + |a_{10}|^2 + |a_{01}|^2 + |a_{00}|^2 = 1. \tag{A3}$$

Equation (A1) may be expressed in terms of the standard Pauli spin vectors by means of the following known relations:

$$\begin{aligned} I_2 &= (|1\rangle\langle 1| + |0\rangle\langle 0|); \quad \tau_1 = (|1\rangle\langle 0| + |0\rangle\langle 1|); \\ \tau_2 &= i(|0\rangle\langle 1| - |1\rangle\langle 0|); \quad \tau_3 = (|1\rangle\langle 1| - |0\rangle\langle 0|). \end{aligned} \tag{A4}$$

Thus

$$\rho(A,B) = \frac{1}{4}\left\{ \begin{aligned} &I_2(A)\otimes I_2(B) + \vec{s}(A)\cdot\vec{\tau}(A)\otimes I_2(B) + I_2(A)\otimes\vec{\tau}(B)\cdot\vec{s}(B) \\ &+\sum_{i,j=1}^{3} C_{ij}(A,B)\tau_i(A)\otimes\tau_j(B) \end{aligned} \right\} \tag{A5}$$

where

$$\begin{aligned} \vec{s}(A) &= (s_1(A), s_2(A), s_3(A)), \\ s_1(A) &= (a_{11}a_{01}^* + a_{11}^*a_{01} + a_{10}a_{00}^* + a_{10}^*a_{00}), \\ s_2(A) &= i(a_{11}a_{01}^* - a_{11}^*a_{01} + a_{10}a_{00}^* - a_{10}^*a_{00}) \\ s_3(A) &= (|a_{11}|^2 - |a_{01}|^2 + |a_{10}|^2 - |a_{00}|^2) \end{aligned} \tag{A6}$$

$$\begin{aligned} \vec{s}(B) &= (s_1(B), s_2(B), s_3(B)), \\ s_1(B) &= (a_{11}a_{10}^* + a_{11}^*a_{10} + a_{01}a_{00}^* + a_{01}^*a_{00}), \\ s_2(B) &= i(a_{11}a_{10}^* - a_{11}^*a_{10} + a_{01}a_{00}^* - a_{01}^*a_{00}) \\ s_3(B) &= (|a_{11}|^2 - |a_{10}|^2 + |a_{01}|^2 - |a_{00}|^2) \end{aligned} \tag{A7}$$

Note that eq.(A7) follows from eq.(A6) by writing $a_{10}$ in place of $a_{01}$. These vectors are sometimes called "polarization" vectors associated with the qubits in analogy with optical polarization vectors.

$$\begin{aligned} C_{11}(A,B) &= \left(a_{11}a_{00}^* + a_{00}a_{11}^* + a_{10}a_{01}^* + a_{01}a_{10}^*\right), \\ C_{12}(A,B) &= i\left(a_{11}a_{00}^* - a_{00}a_{11}^* + a_{10}a_{01}^* - a_{01}a_{10}^*\right), \\ C_{13}(A,B) &= \left(a_{11}a_{01}^* + a_{01}a_{11}^* - a_{10}a_{00}^* - a_{00}a_{10}^*\right), \end{aligned} \tag{A8}$$

$$\begin{aligned} C_{21}(A,B) &= i\left(a_{11}a_{00}^* - a_{00}a_{11}^* - a_{10}a_{01}^* + a_{01}a_{10}^*\right), \\ C_{22}(A,B) &= \left(-a_{11}a_{00}^* - a_{00}a_{11}^* + a_{10}a_{01}^* + a_{01}a_{10}^*\right) \\ C_{23}(A,B) &= i\left(a_{11}a_{01}^* - a_{01}a_{11}^* - a_{10}a_{00}^* + a_{00}a_{10}^*\right) \end{aligned} \tag{A9}$$

$$\begin{aligned} C_{31}(A,B) &= \left(a_{11}a_{10}^* + a_{10}a_{11}^* - a_{10}a_{00}^* - a_{00}a_{10}^*\right) \\ C_{32}(A,B) &= i\left(a_{11}a_{10}^* - a_{10}a_{11}^* - a_{01}a_{00}^* + a_{00}a_{01}^*\right) \\ C_{33}(A,B) &= \left(|a_{11}|^2 - |a_{10}|^2 - |a_{01}|^2 + |a_{00}|^2\right) \end{aligned} \tag{A10}$$

From eq.(A5) we obtain the marginal density matrices

$$\rho(A) \equiv Tr_B\rho(A,B) = \frac{1}{2}\left(I_2(A) + \vec{s}(A)\cdot\vec{\tau}(A)\right), \tag{A11}$$

$$\sigma(B) \equiv Tr_A\rho(A,B) = \frac{1}{2}\left(I_2(B) + \vec{s}(B)\cdot\vec{\tau}(B)\right). \tag{A12}$$

In general these density matrices obey the conditions

$$Tr_A\rho^2(A) = \frac{1}{2}\left[1 + |\vec{s}(A)|^2\right] \leq 1, \tag{A13}$$

the inequality representing mixed and the equality pure state. We observe the following relationship after using the normalization condition in eq.(A3) and the definitions in eqs.(A6,7):

$$1 - |\vec{s}(A)|^2 = 4|a_{11}a_{00} - a_{01}a_{10}|^2 = 1 - |\vec{s}(B)|^2 \tag{A14}$$

Thus the conditions in eq.(A13) is found to be obeyed in general. The concurrence for the generic pure state (A2), $C(AB) = 2\left|a_{11}a_{00} - a_{01}a_{10}\right|$, was given in [8, 9] as well as in [18]. Also eq.(A14) was derived in [18].

We may also mention that a general two-qubit mixed state density matrix can always be expressed in terms of the Pauli spin matrices in the form given in eq.(A5) and the two marginal density matrices are then of the form given in eqs.(A11,12). The relation in eq.(A13) holds in this case as well.

**REFERENCES**

[1] M. A. Nielsen and J. Kempe, Phys. Rev. Lett. **86**, 5184 (2001).

[2] K. G. H. Vollbrecht and M. M. Wolf, e-print: quant-ph/0202058 (2002).

[3] S. Abe and A. K. Rajagopal, Physica A **289**, 157 (2001).

[4] S. Abe, e-print: quant-ph/0104135 (2001) (A revised version will appear in Phys. Rev. A (2002)).

[5] B. M. Terhal and K. G. H. Vollbrecht, Phys. Rev. Lett. **85**, 2625 (2000).

[6] A. O. Pittinger and M. H. Rubin, Phys. Rev. **A62**, 042306 (2000).

[7] K. G. H. Vollbrecht and R. F. Werner, Phys. Rev. **A64**, 062307 (2001).

[8] W.K. Wooters, Phys. Rev. Lett. **80**, 2245 (1998); S. Hill and W.K. Wooters, Phys. Rev. Lett. **78**, 5022 (1997).

[9] V. Coffman, J. Kundu, and W. K. Wootters, Phys. Rev. **A61**, 052306 (2000).

[10] N. J. Cerf and C. Adami, Phys. Rev. Lett. **79**, 5194 (1997).

[11] L. Henderson and V. Vedral, J. Phys. A **34**, 6899 (2001).

[12] H. Olivier and W. H. Zurek, Phys. Rev. Lett. **88**, 017901 (2002).

[13] A. K. Rajagopal and R. W. Rendell, Phys. Rev. **A65** (to appear) (2002)

[14] M. Koashi, V. Buzek, and N. Imoto, Phys. Rev. **A62**, 050302 (2000).

[15] O. Klein, Zeits. F. Phys. **72**, 767 (1931).

[16] A. Sanpera, R. Terrach, and G.Vidal, Phys. Rev. **A58**, 826 (1997); G.Vidal and R.Terrach, Phys. Rev. **A59**, 141 (1998).

[17] W. H. Zurek, e-print: quant-ph/0105127 (2001) for a comprehensive review of his thinking on this subject.

[18] Wang An-Min, Chin. Phys. Lett. **17**, 242 (2000).

Table I

Summary of the examples elucidating Theorems A – D.

| Example | Concurrence C(X,Y) | q-Entropy difference $S_q(A,B)-S_q(A\,or\,B)$ | Quantum Deficit D Mutual Entropy S |
|---|---|---|---|
| (E1) | C(A,B)=2/3 Entangled Mixed | (q=1) –(5/6)ln(5/4) < 0 Entangled (Th. A) (Note: $S_1(A,B)-S_1(B)=0$) | D/ln2= (5ln5-8ln2)/6ln2 =0.6016 S/ln2= $(3\ln 3-2\ln 2)/3\ln 2$ =0.9182 |
| (E2) | C(B,C)=1/3 Entangled Mixed | (q=1) (5/6)ln(5/4) > 0 Entangled (Th. A) | D/ln2=(1/3) S/ln2= $(3\ln 3+8\ln 2-5\ln 5)/3\ln 2$ =0.3817 |
| (E3) | 2/3 Entangled Mixed | 0 for all q Entangled (Th. A) | D/ln2=(2/3) S/ln2=0.9183 |
| (E4) | C(BC)=1 Entangled Pure - Bell state | (q=1) -ln2<0 Entangled (q=1) (Th. B) | D/ln2= 1 S/ln2=2 |
| (E5) | C(A,B)=0 Separable $\rho_{AB}=\rho_A\otimes\sigma_B$ | ln2>0 Separable (Th. D) (Note: $S_1(A,B)-S_1(B)=0$) | D/ln2= 0 S/ln2=0 Classically Uncorrelated |
| (E6) | C(A,B)=0 Separable $\rho_{AB}=\begin{pmatrix}\rho_A\otimes\sigma_B\\+\\\rho_{A'}\otimes\sigma_{B'}\end{pmatrix}\Big/2$ | 0 for all q (Th. D) | D/ln2= 0 S/ln2=1 |

Figure I

Scaled quantum deficit, $D(AB)/\ln 2$ (dashed curve), scaled mutual entropy, $S(A:B)/\ln 2$ ( dot-dashed curve) and concurrence $C(AB)$ (solid curve) for the Werner state $\rho_p$ described in the text.

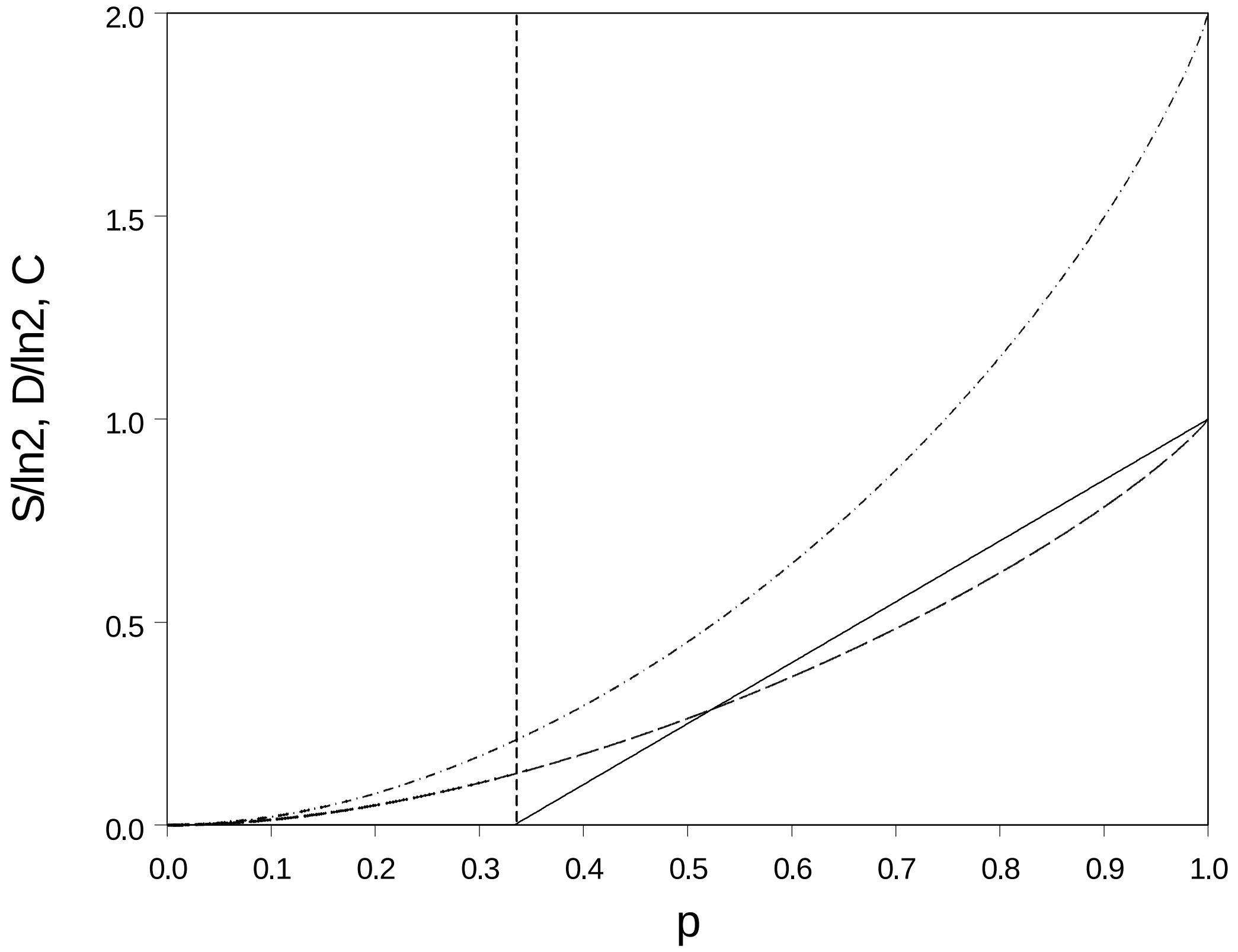